# EFFECT OF SURFACE GROOVES ON THE STATIC FRICTION OF AN ELASTIC SLIDER

R. Capozza[1*], Nicola Pugno[2,3,4]

[1]*International School for Advanced Studies (SISSA), Via Bonomea 265, I-34136 Trieste, Italy*
[2]*Laboratory of Bio-Inspired & Graphene Nanomechanics, Department of Civil, Environmental and Mechanical Engineering, University of Trento, via Mesiano, 77, I-38123 Trento, Italy.*
[3]*Center for Materials and Microsystems, Fondazione Bruno Kessler, Via Sommarive 18, 38123 Povo (Trento).*
[4]*School of Engineering & Materials Science, Queen Mary University of London, Mile End Road, London E1 4NS, UK*

*\*rosario.capozza@gmail.com*

## ABSTRACT

*Numerous studies have proved the usefulness of surface patterning for the modification of tribological performances of sliding contacts. Here we investigate the effects of patterning on the tribological properties of a slider over a solid substrate. We show that, depending on the numerical density of surface grooves, the tribological properties can change significantly. A low density of surface patterning leads to a decrease of static friction force, while a higher density weakens this effect.*

*The decrease is related to a non-uniform distribution of surface stress induced by patterning. We believe these findings and approach to be relevant for technological applications and related optimal design.*

Introduction of specific microtextures onto a sliding surface, comprising flat regions interrupted by cavities, is a valuable approach to modify tribological properties of sliding contacts [1,2]. Small cavities are capable to trap particles, eliminating wear debris from the mating interfaces and increasing the durability of mechanical components.

In the presence of a lubricating medium, the usefulness of these cavities is even more evident. Surface dimples can act as reservoir for lubricant, feeding the lubricating fluid directly between the two contacting surfaces [3,4] and extending the range of hydrodynamic lubrication [5]. This is particularly the case of biological systems where surface textures are very common. Insects are able to efficiently attach to surfaces because their flexible, patterned pads adapt to surface roughness and establish large contact areas.

The attachment pads on the legs of some frogs and bush crickets, display a hexagonal pattern where the volume between hexagons is filled with a secretion liquid [6,7].

Exactly as in the case of tyres, patterning prevents the loss of traction due to building a layer of liquid between the contacting surfaces (hydroplaning) and optimizing the thickness of the fluid film in the presence of secretion fluid [8].

Many experimental and numerical studies demonstrate the effectiveness of surface patterning in reducing friction both at macro and nanoscale [9-13], but the generality of this claim remains controversial. In fact the patterned pads of many insects and reptiles show an extraordinary ability to stick to surfaces with no loss of traction.

In what conditions surface patterning leads to a reduction of friction force?

In this letter we study the effect of surface patterning on dry friction. We consider an elastic block (the slider; of length $l = 10^{-1}$m and cross-sectional area $A_S = 10^{-4}$ m² sliding on an immobile rigid substrate with a rough surface) as illustrated in Fig.1a. The Young's and shear modulus are fixed ($E = 5 \cdot 10^7$Pa and $G = 10^6$Pa respectively) as values typical for rubber.

The elasticity of the slider is taken into account by splitting it into $N_b$ rigid and identical blocks with mass $m_b$, length $l_b = \dfrac{l}{N_b}$ and cross-sectional area $A_S$. The blocks are coupled by a set of springs of stiffness $K_{\text{int}}$. The elasticity of the slider is then $K_{sl} = K_{\text{int}}/(N_b - 1)$. The friction between each block and the substrate is described in terms of $N_s$ elastic interactions representing interfacial asperities [14, 15]. Each asperity is modeled as a surface spring of elastic constant $k_i$ (and average elasticity $k_s = <k_i>$) where $i = 1, 2..., N_s$, and then the shear stiffness $K_{sur}$ of the entire interface [16] is the sum of the contact stiffnesses $k_i$ over all the blocks. The total number of active surface springs is assumed to be proportional to the normal pressure, assumed uniform along the slider. The interaction between blocks is mediated through springs $K_d = K_G/N_b$ (Fig.1a), where $K_G$ represents the shear stiffness of the slider. Numerical values of $K_{sl}$ and $K_G$ are obtained from relations $K_{sl} = \frac{E \cdot A_S}{l}$ and $K_G = \frac{G \cdot S}{l}$, with $S = 10^{-3}$ m², area of top surface of the slider.

When the slider is pulled at constant velocity $V$, at the beginning the surface springs elongate with the velocity of the corresponding block, opposing a force $f_i = k_i l_i(t)$ against the motion, where $l_i(t)$ is the spring length variation. A contact breaks when $f_i$ exceeds a threshold $f_{si}$ and reattaches in an unstressed state, after a delay time $\tau$. Artificial vibrations of the blocks are avoided by introducing a viscous damping force with a coefficient $\gamma$, $f_j = -m_b \dot{x}_j \gamma$, where $x_j$ is the coordinate of the centre of mass of the j-th block. At the beginning of simulations, blocks are randomly displaced from equispaced positions to account for randomness of interface.

The effect of patterning is simulated here by introducing a number of surface grooves, $n_g$, transverse respect to the direction of pulling. A groove of length $l_g$ is obtained by removing surface springs from a group of blocks, as shown in Fig.1b. In all our simulations we keep fixed the total number of surface springs, that corresponds to maintain constant the normal pressure on the slider. By changing the number density of grooves, $\lambda = n_g/l$, we redistribute uniformly surface springs in the contacting regions, the pawls, with length $l_p$. In order to prevent edge effects and minimize stress concentrations due to the pawl's size, grooves are never at the edges of the slider. Unless differently specified, we study the symmetric case $l_p = l_g$ (grooves and pawls of identical size) as shown in Fig.1b in the case of two grooves. This last condition corresponds, in the limit of a large number $n_g$ of grooves, to have half of slider's area covered by pawls, i.e. the packing density is 50%.

Let's assume to drive the slider rigidly from the top surface, as shown in Fig.1. Indicating with $x$ the coordinate of a block of the slider, and with $u(x)$ its displacement respect to the equilibrium, the elastic equation ruling the deformation of the system is [17]:

$$\frac{d^2 u(x)}{dx^2} = \frac{K_{sur} + K_G}{l^2 K_{sl}} (u(x) - a) \qquad (1)$$

where $a = \frac{K_G X_T}{K_{sur} + K_G}$, with $X_T$ position of the top driving surface, that we consider rigid. In the presence of a deformation of amplitude $u_d$ at position $x_d$, a solution of Eq.(1) is

$$u(x) = \begin{cases} u_d\, e^{-\frac{(x-x_d)}{l}\sqrt{\frac{K_{sur}+K_G}{K_{sl}}}} + \frac{K_G X_T}{K_{sur}+K_G} & x > x_d \\ u_d\, e^{\frac{(x-x_d)}{l}\sqrt{\frac{K_{sur}+K_G}{K_{sl}}}} + \frac{K_G X_T}{K_{sur}+K_G} & x < x_d \end{cases} \qquad (2)$$

Relation (2) shows that displacement $u(x)$ is the sum of two terms: the former is a deformation $u_d$ decaying with a length scale $l_c = l\sqrt{\frac{K_{sl}}{K_{sur} + K_G}}$, the latter a contribution due to the displacement $X_T$ of driving top surface. Here we have $l_c = 1.8 \cdot 10^{-3}\,\text{m} \ll l$ that is typical for soft materials as rubber.

Fig.2b is a 2D color map of surface stress as a function of position $x$ along the slider and time (vertical axis). It shows how a deformation decays in the elastic slider. A central block, located at $x=0.05$ m, is displaced at constant velocity respect to its initial position. The surface stress grows until a critical threshold force is reached and a rupture front propagates across the interface at $time \approx 0.25s$. This is shown in Fig.2a reporting a 2D color map of detached springs (red) as a function of position $x$ and time. At $time \approx 0.25$ s in fact a rupture front starts to propagate forward and backward and stops in the middle of the slider. Fig.3 shows the behavior of a deformation

$u_d = 0.36 \mu m$ (corresponding to the red dashed line in Fig.2b at $time \approx 0.1s$) as a function of position $x$ along the slider. It displays a very good agreement with Eq.(2) (red dashed line).

How do grooves influence the rupture dynamics and static friction force?

Fig.4 shows force profiles as a function of time, obtained for a flat and patterned slider when the top surface is displaced at constant velocity *V*. Surface grooves can strongly reduce static friction force, $F_s$, if the density of grooves, $\lambda$, is low (large pawls), while kinetic friction remains almost unchanged. Fig.5 reports the behavior of $F_s$ as a function of patterning density $\lambda$ for packing densities of 33%, 50% and 66%. Irrespective of packing densities, it clearly demonstrates that for large values of density (small pawls), $\lambda > 5$ cm$^{-1}$, static friction is not affected by patterning, while for $\lambda = 0.1$ cm$^{-1}$, $F_s$ drops of about 50%. This intriguing behavior has been also demonstrated recently measuring static friction force between laser surface textured brass samples against sapphire substrates [18].

This simple system suggests why patterning is so common in biological systems and in general in nature: a dense patterning cause no loss in stiction, giving all the numerous advantages of a functionalized surface.

Why does static friction change so much with the density of grooves?

Fig.6a shows 2D color maps of detached surface springs (red) corresponding to $\lambda = 0.2$ cm$^{-1}$ (blue line in Fig.2). In this case two grooves are separated by three pawls. The red horizontal stripe at $time \cong 2.2s$ indicates that surface springs are broken and the slider is set in motion. Fig.6b is the corresponding 2D map of the surface stress evolution. Although the slider is subject to a uniform shear stress, the presence of grooves cause a stress concentration at the edges of pawls, where it is deformed much more easily. Because of that a rupture front is started at values of shear force much lower than the flat case [19]. Once a crack is initiated at the edge, it propagates for the whole length of the pawl and the slider is set in motion.

A groove at the interface represents a 'defect' producing a deformation decaying in the pawl with a length scale $l_c = 1.8 \cdot 10^{-3}$ m $\ll l_p$, where $l_p = 2 \cdot 10^{-2}$ m is the length of a pawl.

Figures 7a and 7b show the evolution of attached surface springs and corresponding stress for an higher density of grooves $\lambda = 2.2$ cm$^{-1}$, corresponding to small pawls and a higher value of static friction force (red line in Fig.2). This time surface springs break almost simultaneously because stress distribution is much more uniform on the pawls. In fact in this case $l_p = 2.2 \cdot 10^{-3}$ m $\cong l_c$ and the deformation at the edge slightly decays in $l_p$.

In order to have a minor reduction of friction force we expect $l_p = \dfrac{l}{2n_g} \leq l_c$, that leads to a critical value of density $\lambda_c = \dfrac{1}{2l_c} = \dfrac{1}{2l}\sqrt{\dfrac{K_G + K_{sur}}{K_{sl}}}$, where $K_{sur}$, shear stiffness of the interface, can be measured experimentally [16]. The dashed vertical line in Fig.4 is an estimation of $\lambda_c$ and demonstrates that static friction remains almost unchanged for $\lambda > \lambda_c$.

**Conclusions**

We have studied here the effect of surface patterning on static friction showing that a low numerical density of surface grooves reduces static friction. This reduction is due to stress concentrations at interface induced by patterning. Above a critical numerical density of grooves, $\lambda_c$, that depends on mechanical properties of the slider, static friction remains unchanged. This suggests why patterning is so common in biological systems and in general in nature: a dense patterning cause no loss in stiction.

**Acknowledgments**

RC is supported by Sinergia Contract CRSII2₁36287/1, and by ERC Advanced Grant 320796 - MODPHYSFRICT. This work is also supported by the COST Action MP1303 "Understanding and Controlling Nano and Mesoscale Friction". NMP is supported by the European Research Council (ERC StG Ideas 2011 BIHSNAM n. 279985 on "Bio-Inspired hierarchical super-nanomaterials", ERC PoC 2013-1 REPLICA2 n. 619448 on "Large-area replication of biological anti-adhesive nanosurfaces", ERC PoC 2013-2 KNOTOUGH n. 632277 on "Super-tough knotted fibres"), by the European Commission under the Graphene Flagship (WP10 "Nanocomposites", n.

**References**

[1] Etsion, I.: Improving Tribological Performance of Mechanical Components by Laser Surface Texturing. Tribol. Lett. **17**, 733 (2004)

[2] Raeymaekers, B., Etsion, I., Talke, F.E.: Enhancing tribological performance of the magnetic tape/guide interface by laser surface texturing. Tribol. Lett. **27**, 89 (2007)

[3] Pettersson, U., Jacobson, S.: Friction and Wear Properties of Micro Textured DLC Coated Surfaces in Boundary Lubricated Sliding. Tribol. Lett. **17**, 553 (2004)

[4] Blatter, A., Maillat, M., Pimenov, S.M., Shafeev, G.A., Simakin, A.V.: Lubricated friction of laser micro-patterned sapphire flats. Tribol. Lett. **4**, 237 (1998)

[5] Mourier, L., Mazuyer, D., Lubrecht, A. A., Donnet, C.: Transient Increase of Film Thickness in Micro-Textured EHL Contacts. Tribol. Int. **39**, 1745 (2006)

[6] Scherge, M., Gorb, S.N.: Biological Micro- and Nanotribology: Nature's Solutions. Springer, Berlin, Germany, (2001)

[7] Murarash, B., Itovich Y., Varenberg, M.: Tuning elastomer friction by hexagonal surface patterning. Soft Matter, **7**, 5553 (2011)

[8] Federle, W., Barnes, W.J.P., Baumgartner, W., Drechsler, P., Smith, J.M.: Wet but not slippery: boundary friction in tree frog adhesive toe pads. J. R. Soc. Interface, **3**, 689 (2006)

[9] Borghi, A., Gualtieri, E., Marchetto, D., Moretti, L., Valeri, S.: Tribological Effects of Surface Texturing on Nitriding Steel for High-Performance Engine Applications. Wear, **265**, 1046 (2008)

[10] Gualtieri, E., Pugno, N., Rota, A., Spagni, A., Lepore, E., Valeri, S: Role of Roughness Parameters on the Tribology of Randomly Nano-Textured Silicon Surface. J. Nanosci. Nanotechnol. **11**, 9244 (2011)

[11] Pugno, N.: Nanotribology of Spiderman. In: Bellucci, S. (eds.) Physical Properties of Ceramic and Carbon Nanoscale Structures, pp 111-136. Springer Berlin, Heidelberg (2011)

[12] Gualtieri, E., Borghi, A., Calabri, L., Pugno, N., Valeri, S.: Increasing nanohardness and reducing friction of nitride steel by laser surface texturing. Tribol. Int., **42**, 699 (2009)

[13] Capozza, R., Fasolino, A., Ferrario, M., Vanossi, A.: Lubricated friction on nanopatterned surfaces via molecular dynamics simulations. Phys. Rev. B **77**, 235432 (2008)

[14] Braun, O.M., Barel, I., Urbakh, M.: Dynamics of Transition from Static to Kinetic Friction. Phys. Rev. Lett. **103**, 194301 (2009)

[15] Capozza, R., Rubinstein, S.M., Barel, I., Urbakh, M., Fineberg, J.: Stabilizing Stick-Slip Friction. Phys. Rev. Lett. **107**, 024301 (2011)

[16] Berthoud, P., Baumberger, T.: Shear stiffness of a solid–solid multicontact interface. Proc. R. Soc. Lond. A **454**, 1615 (1998)

[17] Landau, L.D., Lifshitz, E.M.: Theory of Elasticity, Course of Theoretical Physics Vol. 7. Pergamon, New York (1986)

[18] Greiner, C., Schafer, M., Pop, U., Gumbsch, P.: Contact splitting and the effect of dimple depth on static friction of textured surfaces. Appl. Mater. Interfaces **6**, 7986 (2014)

[19] Capozza, R., Urbakh, M.: Static friction and the dynamics of interfacial rupture. Phys. Rev. B **86**, 085430 (2012)

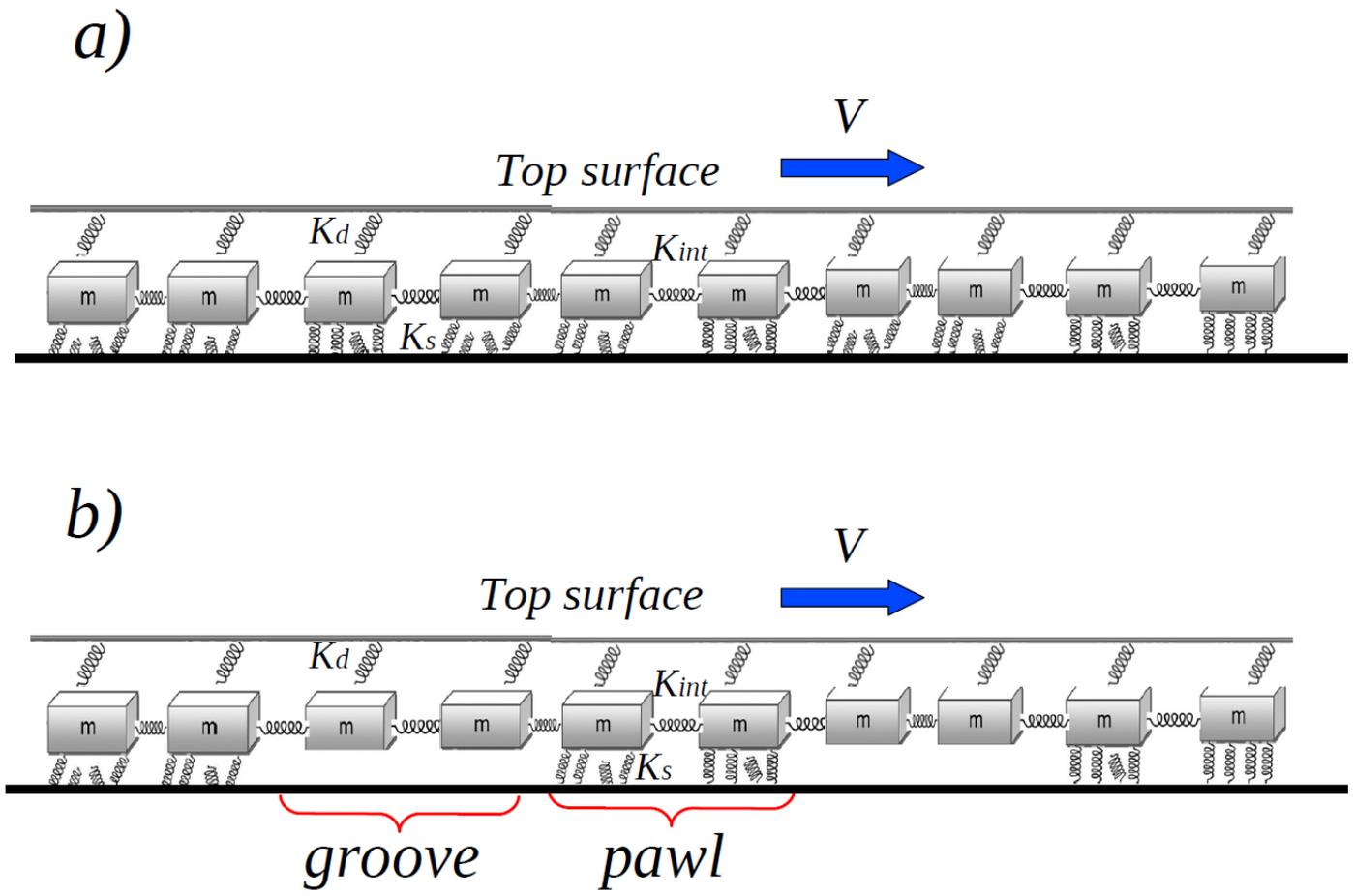

Fig.1 Schematic sketch of the two studied setups showing a flat (a) and patterned (b) slider. Parameter values used in the model: $m_b = 1.59 \cdot 10^{-4}$ Kg, $l_b = 1.6 \cdot 10^{-4}$ m, $N_b = 630$, $K_{sur} = 10^7$ N/m, $N_s = 20$, $f_{si} = 2.4 \cdot 10^{-3}$ N, $\tau = 5 \cdot 10^{-3}$ s, $\gamma = 2.5 \cdot 10^2$ s$^{-1}$

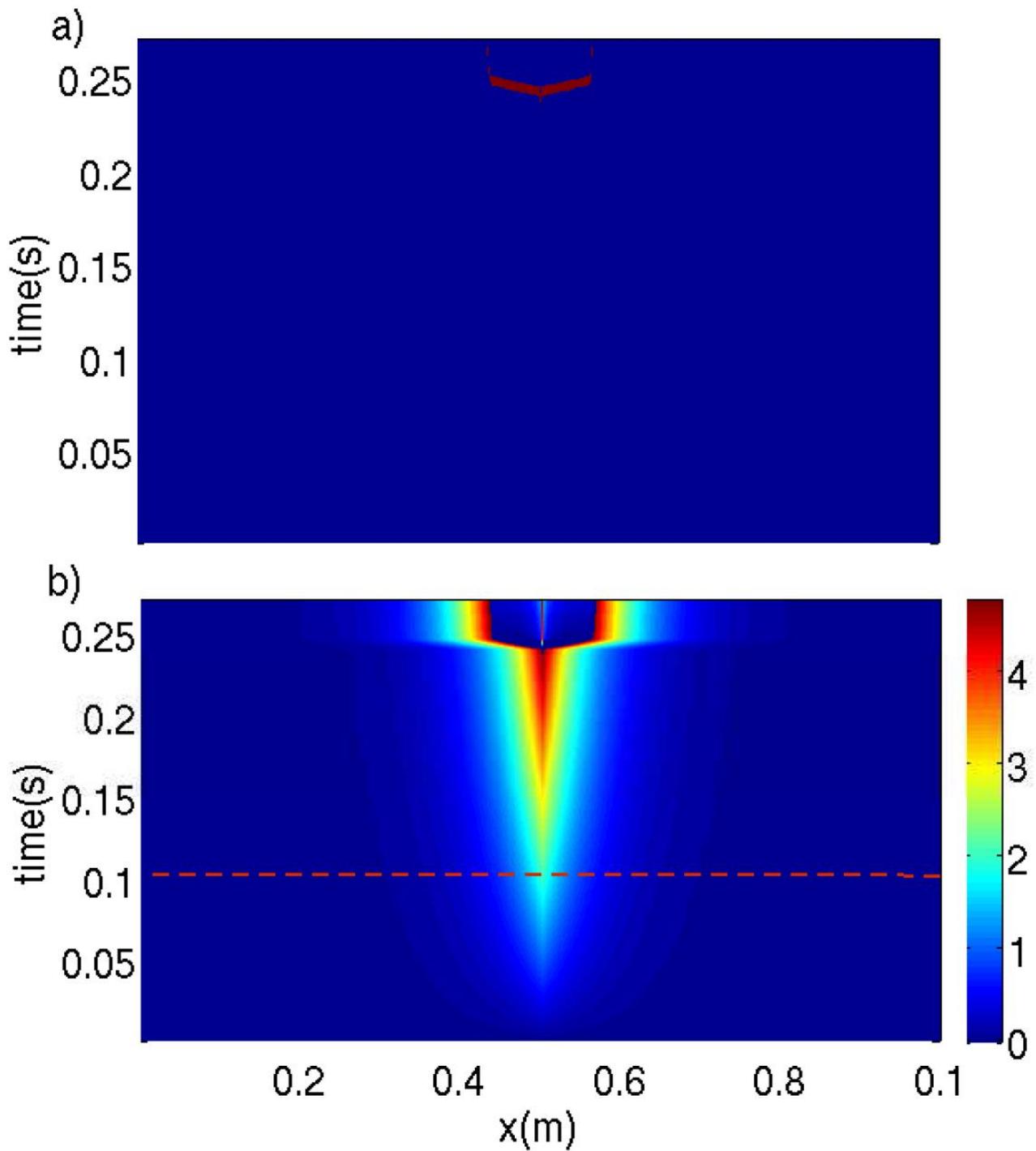

Fig.2 a) Colour map showing the attached (blue) and detached (red) surface contacts as a functions of coordinate *x* along the slider and time. b) 2D maps of surface stress evolution obtained by displacing the central block at constant velocity. Hotter (colder) colours indicate regions of higher (lower) stress. The bar to the right of the map set up a correspondence between the colours and the values of the force in Newton.

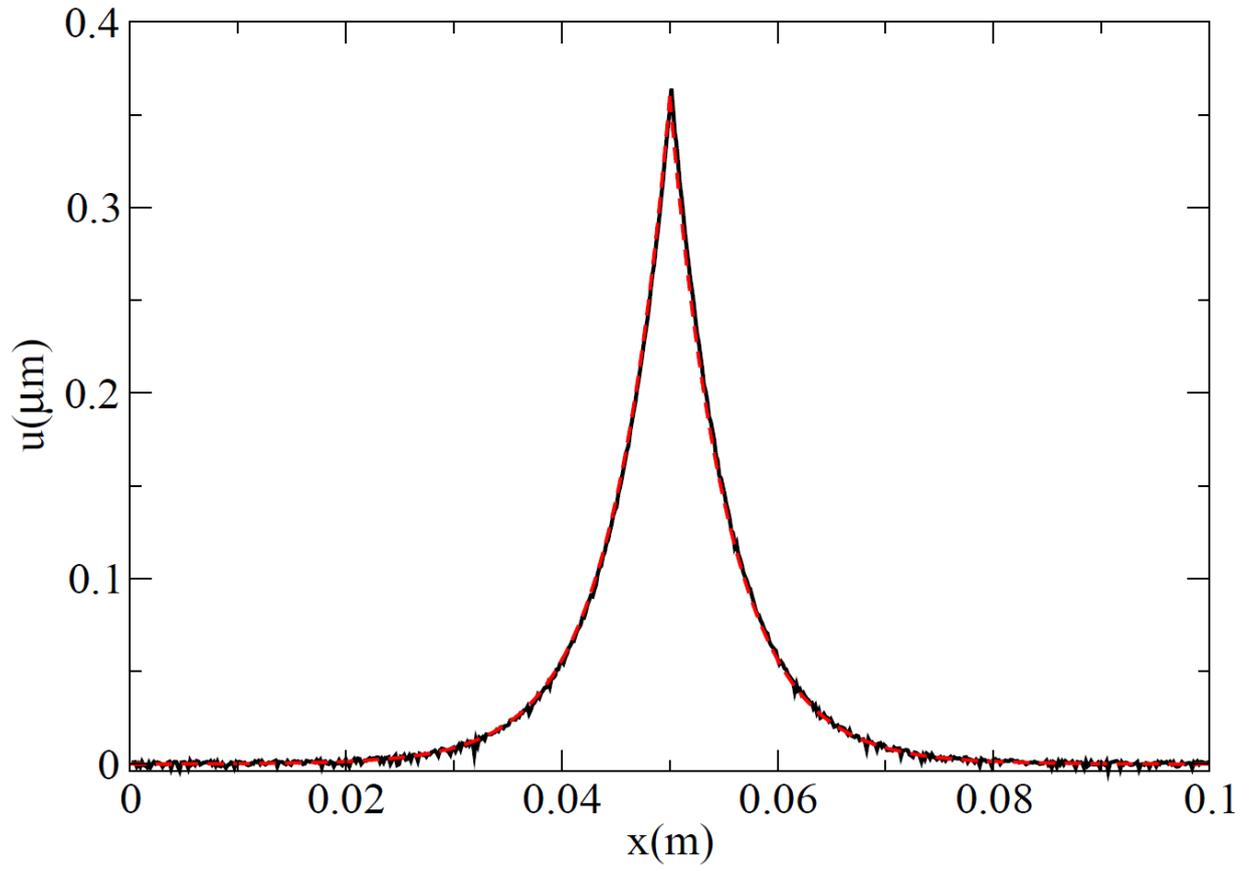

Fig.3 Behavior of a deformation located in the center of the slider, corresponding to the red dashed line in Fig.2b at $time \approx 0.1s$, as a function of position $x$ along the slider (black line). It displays a very good agreement with Eq.(2) (red dashed line).

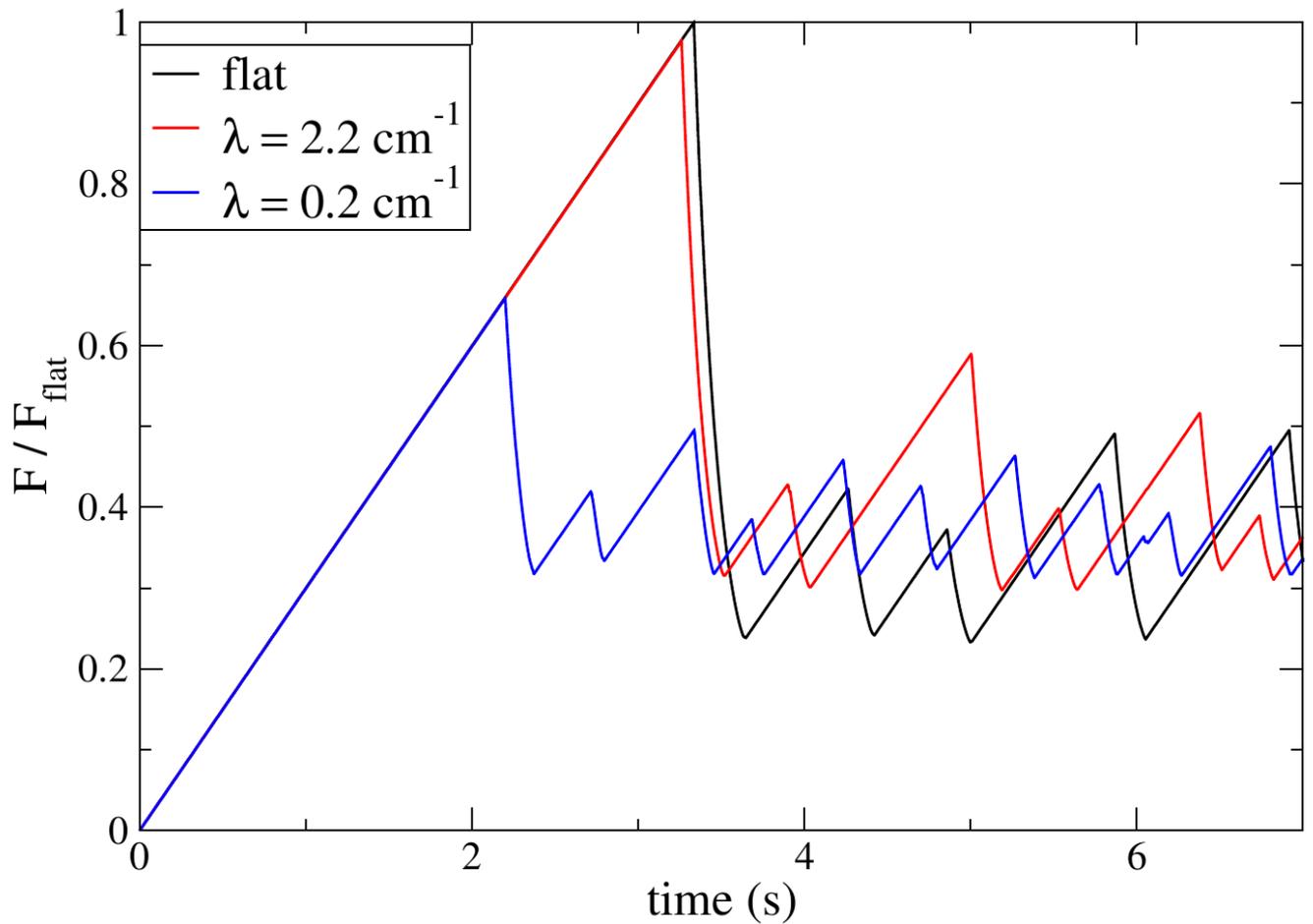

Fig.4 Normalized shear force F/$F_{flat}$ as a function of time for a flat (black) and patterned (red and blue) slider. $F_{flat}= 30$ N is the static friction force obtained in the flat case. Here different colors refer to values of patterning density $\lambda$ indicated in figure. Surface patterning reduces static friction force despite the total number of surface springs (i.e. the real area of contact) remains constant.

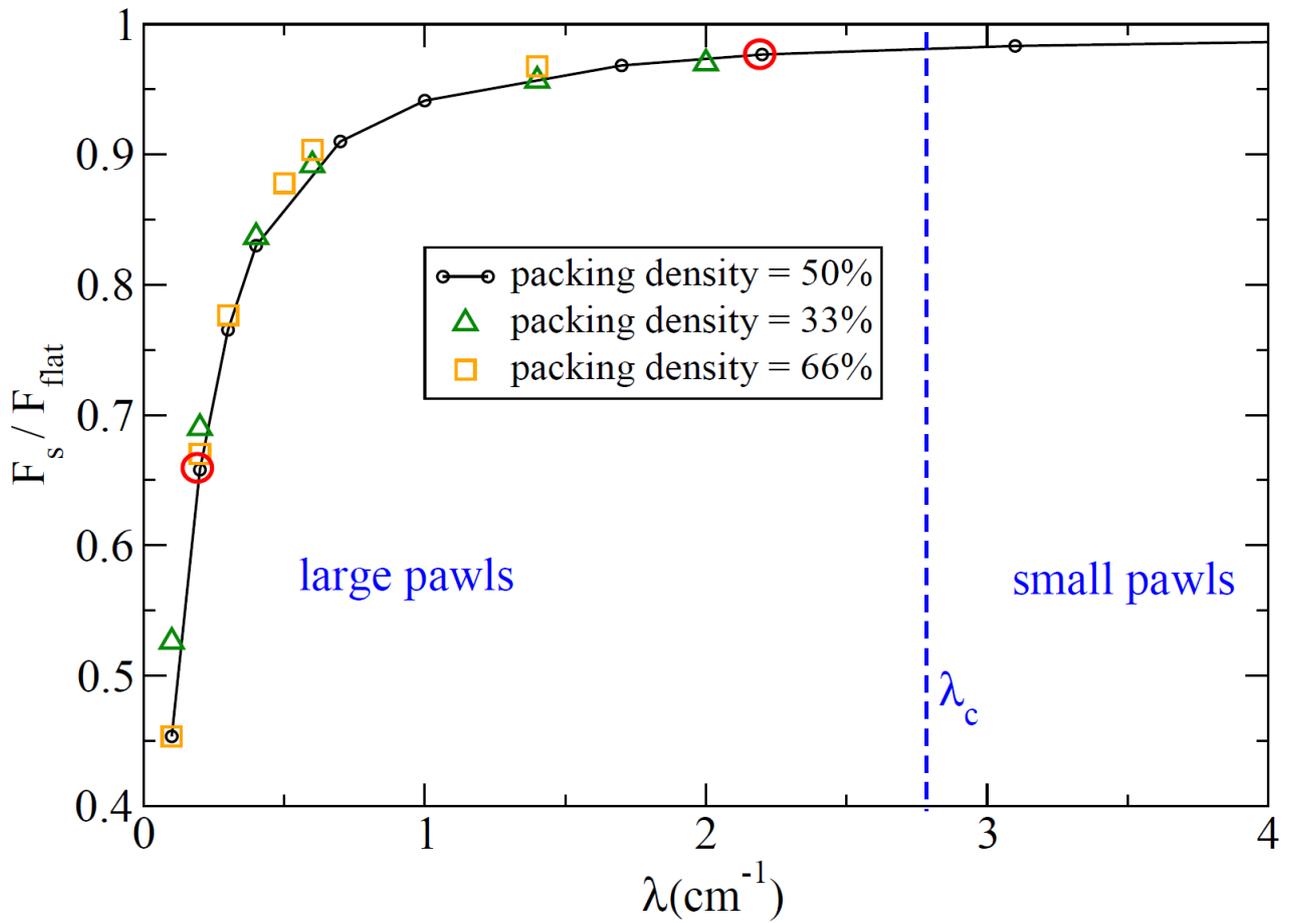

Fig.5 Normalized static friction force $F_s/F_{flat}$ as a function of patterning density, $\lambda$, for indicated values of packing densities. The blue vertical line indicates the value of a critical density of grooves, $\lambda_c$ above which static friction becomes independent on patterning. Red circles indicate values of $F_s$ obtained from force profiles in Fig.2.

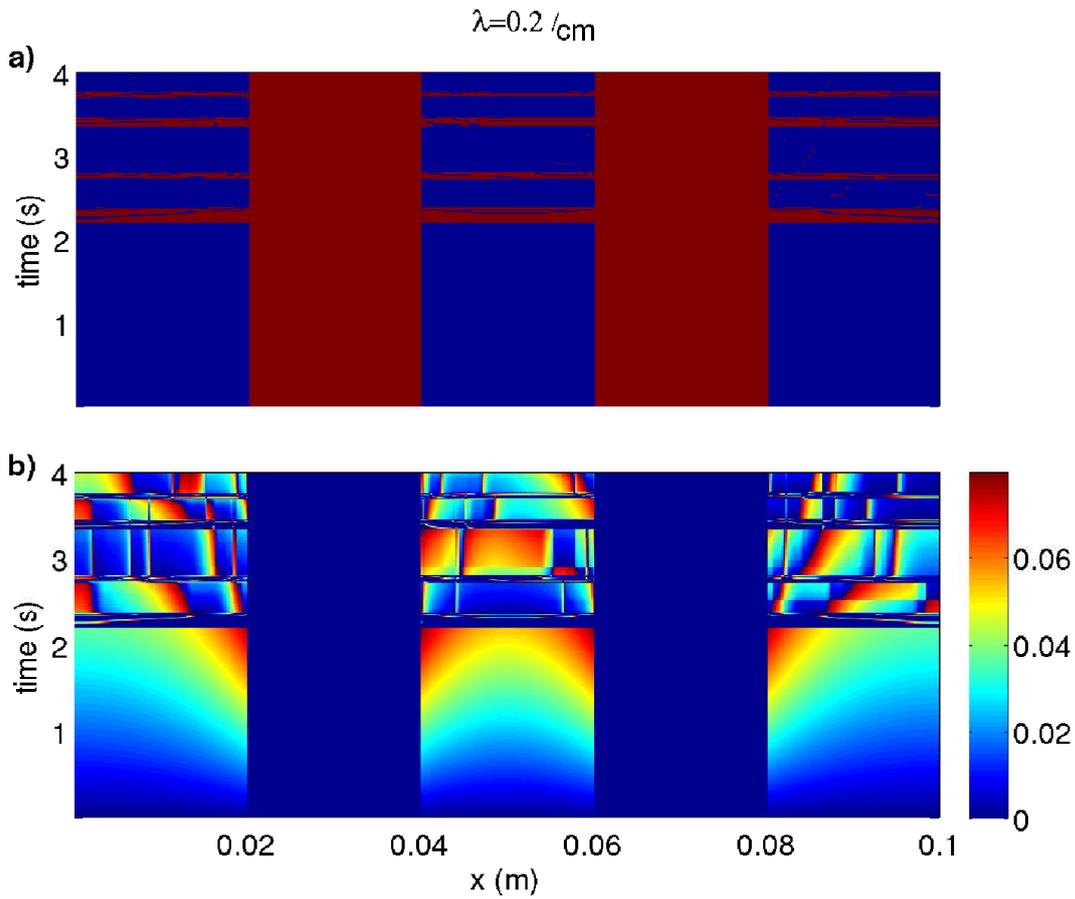

Fig.6 a) Colour map showing the attached (blue) and detached (red) surface contacts as a functions of coordinate *x* along the slider and time. b) 2D maps of surface stress evolution. Hotter (colder) colours indicate regions of higher (lower) stress. The bar to the right of the map set up a correspondence between the colours and the values of the force in Newton. Here $\lambda = 0.2 \text{cm}^{-1}$.

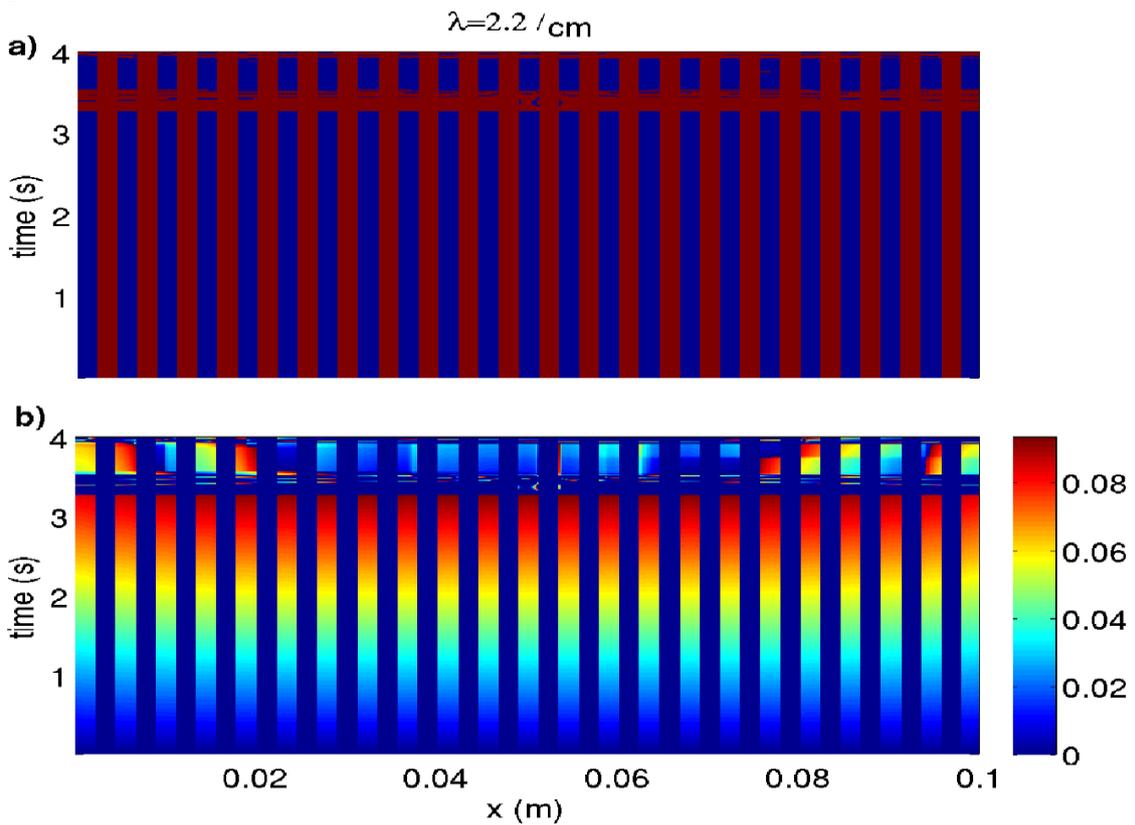

Fig.7 a) Colour map showing the attached (blue) and detached (red) surface contacts as a functions of coordinate $x$ along the slider and time. b) 2D maps of surface stress evolution. Hotter (colder) colours indicate regions of higher (lower) stress. The bar to the right of the map set up a correspondence between the colours and the values of the force in Newton. Here $\lambda = 2.2 cm^{-1}$.